\documentclass[11pt]{article}
\usepackage{amsmath}    
\usepackage{amssymb}
\usepackage{graphicx}
\usepackage{amsfonts}   
\usepackage{latexsym}   
\usepackage{slashed}
\usepackage{yfonts} 
\usepackage{verbatim}
\usepackage{wrapfig}

\let\non\nonumber

\global\newcount\itemno \global\itemno=0

\def\itemaut#1{\global\advance\itemno by1\noindent\item{\the\itemno.}#1}

\newif{\ifeq}           
\eqtrue                 

\newcommand{\be}{\begin{equation}}
\newcommand{\ee}{\end{equation}}
\newcommand{\bes}{\begin{equation*}}
\newcommand{\ees}{\end{equation*}}
\newcommand{\bea}{\begin{eqnarray}}
\newcommand{\eea}{\end{eqnarray}}
\newcommand{\bean}{\begin{eqnarray*}}
\newcommand{\eean}{\end{eqnarray*}}

\def\({\left(}
\def\){\right)}
\def\[{\left[}
\def\]{\right]}

\def\frac#1#2{{#1 \over #2}}

\newcommand{\half}{\frac{1}{2}}

\renewcommand{\a}{\alpha}
\renewcommand{\b}{\beta}

\renewcommand{\r}{\rho}
\newcommand{\s}{\sigma}

\newcommand{\m}{\mu}

\renewcommand{\t}{\theta}

\newcommand{\IP}{{\mathbb P}}

\newcommand{\lsim}{\,\raise.3ex\hbox{$<$\kern-.75em\lower1ex\hbox{$\sim$}}\,}
\newcommand{\gsim}{\,\raise.3ex\hbox{$>$\kern-.75em\lower1ex\hbox{$\sim$}}\,}

\def\p{\partial}

\def\II{\relax{I\kern-.10em I}}

\usepackage{amssymb}

\def\cg{{\rm g}}
\def\k{\kappa}

\numberwithin{equation}{section}

\begin{document}

\begin{titlepage}
\begin{flushright}COLO-HEP-548 \end{flushright}
\vskip 1in

\begin{center}
{\Large{Charged Schr\"odinger Black Holes}}
\vskip 0.5in Allan Adams$^{1}$, Charles Max Brown$^{2}$, Oliver DeWolfe$^{2}$, Christopher Rosen$^{2}$
\vskip 0.4in {\it $^{1}$ Center for Theoretical Physics\\  Massachusetts Institute of Technology\\ Cambridge, MA  02139, USA}
\vskip 0.2in {\it $^{2}$ Department of Physics 390 UCB \\ University of Colorado \\ Boulder, CO 80309, USA}
\end{center}
\vskip 0.5in

\begin{abstract}\noindent
We construct charged and rotating asymptotically Schr\"odinger black hole solutions of IIB supergravity.
We begin by obtaining a closed-form expression for the null Melvin twist of a broad class of type IIB backgrounds, including  solutions of minimal five-dimensional gauged supergravity, and identify the resulting five-dimensional effective action.  
We use these results to demonstrate that the 
near-horizon physics and thermodynamics of asymptotically Schr\"odinger black holes obtained in this way are essentially inherited from their $AdS$ progenitors, and verify that they admit zero-temperature extremal limits with $AdS_{2}$ near-horizon geometries.  Notably, the $AdS_2$ radius is parametrically larger than that of the asymptotic Schr\"odinger space.

\end{abstract}

\end{titlepage}

\section{Introduction}

A series of beautiful experiments on cold atoms at unitarity \cite{zwerger} and other non-relativistic critical systems has led to intense study of non-relativistic conformal field theories (NRCFTs) \cite{NishidaSon}.  Since these systems are generally strongly non-perturbative, making progress with traditional tools has proven difficult.  In an attempt to provide a strong-coupling expansion, Son \cite{Son} and Balasubramanian and McGreevy \cite{BM} proposed a new application of gauge-gravity duality to NRCFTs in which the usual AdS$_{d+1}$ space, whose isometry group is the relativistic conformal group  in $d$ space-time dimensions,  is replaced by the Sch$_{d+2}$ geometry, $$
ds^{2} = -{dt^{2}\over r^{4}} + {2dt d\xi +d\vec{x}^{2}+dr^{2}\over r^{2}}\,,
$$
whose isometry group is the non-relativistic conformal group in $d$ space-time dimensions -- the so-called Schr\"odinger group.  In this construction, $\xi$ is a compact null circle, with the particle number of the NRCFT identified with momentum along this DLCQ'd direction, $N=i\p_{\xi}$.  Using a simple generalization of the usual AdS/CFT dictionary, one can show that this gravitational dual does indeed reproduce the correlation functions of a NRCFT at zero temperature and chemical potential.

Of course, a non-relativistic system at zero temperature and zero density is a somewhat degenerate system.  Meanwhile, the Sch$_{d+2}$ geometry above has a null singularity in the IR region near $r\to\infty$.  To improve the situation, one can turn on a non-zero temperature and chemical potential.  In the gravity solution, this corresponds to hiding the erstwhile null singularity behind a warm fuzzy black hole horizon at finite $r$ \cite{ABM,MMT,HRR}.  Using the previously-conjectured dictionary, these gravity systems generate boundary correlation functions which transform like those of an NRCFT at finite temperature and density, which pass some rather non-trivial checks \cite{FuertesMoroz}, and which lead to the prediction that the dual NRCFT has a viscosity to entropy ratio which saturates the KSS bound \cite{KSS}.  The evidence is thus strong that these systems describe some NRCFT living on the boundary.

Our main interest, however, is not in the symmetric phase of these systems, but rather in superfluid phases in which the particle-number $U(1)$ is spontaneously broken by the particle condensate.  In the relativistic context, such condensates have been studied in some detail by Gubser \cite{Gubser} and by Hartnoll, Herzog and Horowitz \cite{HHH} and others, where charged black holes were shown to be unstable to the spontaneous emission of charge into a trapped layer above their horizons -- a near-horizon superfluid.  Rotation is readily incorporated by studying charged rotating black holes \cite{Sonner}.  Relatedly, in the context of fermions, the study of fermi condensates involves a very similar structure, with a sharp fermi surface appearing in the zero temperature extremal limit of a charged black hole \cite{LMV}.  It would be very interesting to extend these construction to the non-relativistic regime.

We are thus led to search for more general charged rotating asymptotically Schr\"odinger black holes.  In this paper, we will construct precisely such black hole solutions in IIB supergravity.  As in the original example, we will derive these solutions from charged, rotating AdS solutions by the application of a solution generating technique, the null Melvin twist.  Also as in the original example, these black holes will inherit many of their properties from AdS space.  However, they will enjoy a number of novel features.  First, these solutions will boast extremal limits with AdS$_{2}$ near-horizon geometries, a very powerful constraint on the IR physics.  Secondly, unlike the AdS case, the radius of curvature of the near-horizon region will be independent of the radius of curvature of the asymptotic geometry, but will instead be controlled by the NR density of the boundary CFT.  Third, these solutions will allow us to explicitly break rotation invariance in the boundary solution with either rotation or additional fields.  

In constructing these solutions explicitly, several bits of technology will be quite useful.  First, we obtain a closed-form expression for a general melvinization, including RR forms, which simplifies our computations.  Secondly, we determine the dimensionally reduced five-dimensional action associated to the Melvinized solutions.  The Melvin map on five dimensional solutions is shown to slightly modify the metric, preserve the initial massless gauge field, and introduce a massive gauge field corresponding to the Killing vector in the direction of the number operator isometry.

This paper is organized as follows.  In Section 2, we present the closed-form expression for the null Melvin twist, restrict the general form to solutions of 5d gauged supergravity, and discuss the 5d effective action after Melvinization.  Section 3 addresses general properties of black holes which arise from Melvinization, including their thermodynamics and near-horizon geometries.  Section 4 works out two examples, the charged RN-Sch$_{5}$ black hole and the charged-rotating Kerr-Newman-Sch$_{5}$ black hole, including a brief discussion of their thermodynamics and near horizon geometries.  We conclude in Section 5.

\section{The Null Melvin Twist}
\label{MelvinSec}

The Null Melvin Twist \cite{Alishahiha:2003ru,Gimon:2003xk}\ is a simple solution generating technique for IIB supergravity.  The idea is to act with a series of symmetry operations which preserve, point by point, the satisfaction of the equations of motion, but which do not necessarily respect global properties of the solution.  The result is a new and generally inequivalent solution which, however, inherits some of the structure of the original background.  This process of ``Melvinization" has been used successfully to construct (up to now uncharged, unrotating) black holes with asymptotic Schr\"odinger symmetry from AdS black holes in type IIB supergravity.  With an eye toward extremal solutions with Schr\"odinger asymptotics, we would like to apply this technique to AdS black holes carrying charge and rotation.

For this technique to work, the initial solution must admit one timelike and two spacelike isometries, $\p_{\tau}$, $\p_{y}$ and $\p_{\chi}$.  One then performs a boost of rapidity $\gamma$ in the $\tau$-$y$ plane followed by a T-duality along $y$, and then twists $d\chi \to d\chi + \a dy$; this is followed by another T-duality along $y$ and a boost of rapidity $-\gamma$ in the $\tau$-$y$ plane.  Finally one takes a double-scaling limit with
\begin{eqnarray}
\alpha \to 0 \,, \quad \gamma \to \infty \,, \quad \quad \beta \equiv \alpha \cosh \gamma \;\; {\rm fixed}\,.
\end{eqnarray}
The ``Melvinized" solution is characterized by the new constant $\beta$; the original solution is recovered in the limit $\beta \to 0$.  By construction, the end result is again a solution of 10d IIB supergravity.

While the procedure is completely straightforward in principle, intermediate steps can be messy.  For simplicity, we begin this section with a closed-form expression for the Melvinization of a broad class of initial solutions, then specialize to 10d solutions corresponding to general AdS$_{5}$ black holes. We use the result as an ansatz for a dimensional redution to five dimensions and present the reduced 5d action.  We will use these results to discuss the thermodynamics of Melvinized black holes in the next section, then discuss some simple examples in the subsequent section.

\subsection{General Melvin map}

We consider a type IIB supergravity background with the requisite three isometries, a self-dual five form $F_5$, constant dilaton and all other fields vanishing.  These assumptions can be relaxed straightforwardly, but will suffice for our purposes.  

Letting $a, b$ run over the isometry directions $\tau$, $y$, $\chi$, with $x^{i=3\dots 9}$ the remaining seven coordinates, the initial background may be written as,
\def\otimes{}
\bea
ds^2 &=& \cg_{ab} \,e^{a}\!\otimes e^{b} \,+ {\cal G}_{ij}\, dx^i\!\otimes dx^j \,, \\
F_5 = * F_5 &=& F_5^0 \,,  \\
 \Phi&=& \Phi_0 \,,
\eea
where the one-forms $e^{a}$, given by
\begin{eqnarray}
e^{\tau} \equiv d\tau\!+\!A^{\tau}_{~i}(x)dx^{i}, \quad \quad
e^{y} \equiv dy\!+\!A^{y}_{~i}(x)dx^{i}, \quad \quad
e^{\chi} \equiv d\chi\!+\!A^{\chi}_{~i}(x)dx^{i}\,,
\label{OneForms}
\end{eqnarray}
are defined so that all cross terms between the three-dimensional part of the metric spanned by $\tau$, $y$, $\chi$ and the other seven directions $x^i$ are absorbed into it; however, they do not contain the cross terms between $\tau$, $y$ and $\chi$ themselves, which are realized by off-diagonal elements in $\cg_{ab}$.  

In writing down the Melvinized solution, it is useful to decompose the 5-form $F_5^0$ into all possible tensors relative to $e^\tau$, $e^y$ and $e^\chi$ and single out two terms in particular:
\begin{eqnarray}
\label{FiveDecomp}
F_5^0 \equiv  e^\tau\wedge e^y \wedge e^\chi \wedge a_2 + {1 \over 2} (e^y -e^\tau) \wedge e^\chi \wedge b_3 + \ldots \,,
\end{eqnarray}
where $a_2$ and $b_3$ are a 2-form and a 3-form with indices in the $x^i$ directions; all other possible terms consistent with self-duality can be present as well, but these two play a special role in the results.

We now perform the Melvinization as described.  A compact and convenient presentation of the T-duality rules is given in appendix~\ref{TDualApp}.
After some character-building labor, we find the result
\bea
{ds^2}' &=& {\beta^2 ||\cg_{ab}||\over K} (e^{\tau}\!+e^{y})^{\otimes 2} + {1 \over K}  \, \cg_{ab} \,e^{a}\!\otimes e^{b}  + {\cal G}_{ij}\, dx^i\otimes dx^j\,,\\
{B_{2}}'&=&B'_{a b} \,e^{a}\!\wedge e^{b} \,, \\
{F_3}' &=& \beta \left[ (e^\tau + e^y) \wedge a_2 + b_3 \right] \,, \\
{F_5}' &=& F_5^0 + B_2' \wedge F_3'\,, \\
{\Phi}' &=& \Phi_0 - {1 \over 2} \log K  \,,
\eea
where $||\cg_{ab}||$ is the determinant of the initial metric in the $\tau$-$y$-$\chi$ directions, and
\begin{eqnarray}
K &=& 1 + \beta^2 \[\cg_{\chi\chi} (\cg_{\tau\tau}+ \cg_{yy} - 2 \cg_{\tau y}) -(\cg_{y\chi} - \cg_{\tau\chi})^2\] \,, \\
B_{\tau y}' &=& {\b \over K} \[\cg_{\tau\chi} (\cg_{yy}- \cg_{\tau y}) + \cg_{y\chi}(\cg_{\tau\tau} - \cg_{\tau y})\]\,, \\
B_{\tau \chi}' &=&{\b \over K} \[ \cg_{\tau\chi} (\cg_{y\chi} - \cg_{\tau\chi}) + \cg_{\chi\chi} (\cg_{\tau\tau} - \cg_{\tau y})\]\,, \\
B_{y\chi}' &=&{\b \over K} \[\cg_{y\chi}( \cg_{y\chi} - \cg_{\tau\chi}) + \cg_{\chi\chi} (\cg_{\tau y} - \cg_{yy}) \]\,.
\end{eqnarray}
Note that the only change in the metric is the squashing of the $(\tau,y,\chi)$ part by $1/ K$ and the addition of a single new term in the $(\tau + y)$-direction. In addition a B-field is generated, as well as a varying dilaton. 

 We also see that the RR flux now includes not only a 5-form but also, in general, a 3-form.
The RR 3-form is non-vanishing only if the five-form terms written explicitly in (\ref{FiveDecomp}) are nonzero.  For all the explicit examples considered in this note, $\tau$ and $y$ will be part of a noncompact five-dimensional geometry, $M$, and $\chi$  will be on a compact Sasaki-Einstein space; when these two factors of the geometry are orthogonal in the metric and $F_5^0 = (1 + *) {\rm vol}_M$, as is the case for the uncharged, non-rotating black hole examples of \cite{HRR, MMT, ABM}, it is easy to see $F_3$ vanishes.   In general, however, $F_3$ is nonzero, as will be the case for the charged examples we study in the following sections.

\subsection{Light Cone Coordinates}
The form of the solution simplifies considerably if we rotate initial and final solutions to a light-cone frame defined by,
$$
t = \b(\tau + y)\,, ~~~~~~~~ \xi={1\over 2\b}(y-\tau)\,.
$$
In the asymptotically Schr\"odinger examples of interest in the following sections, this is also the physically useful choice, as $\p_t$ becomes the canonically normalized generator of time translations in the asymptotic region.  Note that this transformation is unimodular, so we don't have to worry about Jacobians.

In light cone frame, the new term in the metric becomes diagonal, and the Melvinized solution takes the simpler form,
\bea
{ds^2}' &=& {||\cg_{ab}||\over K} e^{t}\otimes e^{t} + {1 \over K}  \, \cg_{ab} \,e^{a}\!\otimes e^{b} + {\cal G}_{ij}\, dx^i\otimes dx^j \,, \\
B_{2}'&=& {\sqrt{|\cg|}\over K} *_{\cg} \!e^{t} \,, \\
F_{3}' &=& i_{\chi}i_{\xi}F^0_{5}\,, \\
F_{5}' &=& F^0_{5}+ B_{2}'\wedge F_{3}' \,,\\
\Phi'&=& \Phi_0 - {1 \over 2} \log K \,,
\eea
where $i$ is the inclusion acting as $i_X (e^{X} \wedge Y) \equiv Y$, and
$$
K = 1 +\[g_{\chi\chi}g_{\xi\xi} -g_{\xi\chi}^{2}\] \,.
$$
The components of $B$ take a simple form in terms of minors,
$$
B_{t\xi} = {1\over K} (g_{\xi\xi}g_{t\chi} -  g_{t\xi}g_{\xi\chi}) \,,
~~~~
B_{t \chi} = {1\over K}(g_{\chi\xi}g_{t\chi} - g_{t\xi}g_{\chi\chi}) \,,
~~~~
B_{\xi\chi} = {1\over K}( g_{\xi\chi}^{2} -g_{\xi\xi}g_{\chi\chi}) \,.
$$
Note that all factors of $\beta$ have been absorbed by the change of coordinates.
%
%

\subsection{Melvinizing Solutions of 5d Gauged Supergravity}
\label{CompactSec}

Melvin maps ten-dimensional solutions to ten-dimensional solutions.  In using Melvin to build Schr\"odinger black holes, we will  generally begin with a five-dimensional AdS black hole, lift it to ten dimensions, Melvinize, and then reduce back to five dimensions.   It is therefore convenient to write down a formula circumventing the side-trip to ten dimensions, and simply mapping one five-dimensional geometry into another.  In this subsection we choose a 10d ansatz corresponding to 5d gauged supergravity and construct a five to five map.

Suppose our initial metric solution takes the Kaluza-Klein form \cite{Chamblin:1999tk, Cvetic:1999xp},
\bea
\label{BHTen}
ds^2_{10} &=& ds^2_5(M) + ds^2(X) +   (\eta + A_q)^2 \,,
\eea
where $M$ is a five-dimensional Lorentzian spacetime on which the isometries $\p_t$ and $\p_\xi$ act, $X$ is a 
compact K\"ahler-Einstein 4-manifold, $\eta \equiv d\chi + {\cal A}$ defines a 
Sasaki-Einstein
fibration $Y$ over $X$, and $A_{q}$  is a 1-form on $M$.  For example if $X = \IP^2$, $\eta$ is then the Hopf fiber on $S^5$; see appendix~\ref{CoordApp} for more details. 
 We can always cast the 5d metric in the form,
\begin{eqnarray}
\label{BHFive}
ds^2_{5}(M) = G_{\a\b}\, e^{\a} e^{\b} + G_{mn} dx^m dx^n \, ,
\end{eqnarray}
with $\a,\b$ running over only $t,\xi$ and $m,n$ over the remaining three spatial dimensions, where $e^t \equiv dt + A^t_m dx^m$ and $e^\xi \equiv d\xi + A^\xi_m dx^m$ contain the cross terms between the $t,\xi$ directions and the other three, but not each other.   (One may equally well use the coordinates $\tau$, $y$, for which analogous statements hold.) Additionally, the 1-form $A_q$ can be written as
\begin{eqnarray}
A_q = A_{q(\alpha)} e^\alpha + A_{q (m)} dx^m \,.
\end{eqnarray}
 We assume as before a constant dilaton, self-dual five-form $F_5^0$ and no other fields.
 The 5-form that supports the metric (\ref{BHTen}) has the form \cite{Chamblin:1999tk, Cvetic:1999xp, MMT}
\begin{eqnarray}
\label{F5Ten}
F_5^0 &=&- 4 {\rm vol}_M + {1 \over 2} (\eta + A_q) \wedge d{\cal A} \wedge d{\cal A} \\
&&+ {1 \over 2} *_5 F_q \wedge d {\cal A}  - {1 \over 2} F_q \wedge  (\eta + A_q) \wedge d{\cal A} \,, \nonumber
\end{eqnarray}
where each line is self-dual by itself; the top line contains the volume form on $M$ and its Hodge dual, and the second line contains a factor of $F_q \equiv d A_q$.  The Bianchi identity $dF_5^0 = 0$ implies
\begin{eqnarray}
\label{FqEqn}
d *_5 F_q= F_q \wedge F_q \,,
\end{eqnarray}
which must be satisfied by any initial solution.
Solutions of the form (\ref{BHTen}), (\ref{F5Ten}) include the various charged, rotating $AdS_{5}$ black holes which we will study.  The dimensional reduction of this ansatz is precisely minimal gauged supergravity in five dimensions,
\begin{eqnarray}
\label{FiveActionTrunc}
2 \kappa_5^2\,S_5 = \int {\rm vol}_M (R^{(5)}  + 12 )
 - {3 \over 2} \int F_q \wedge *_5 F_q+ \int A_q \wedge F_q \wedge F_q \,, 
 \end{eqnarray}
 where the canonically normalized gauge field is $A \equiv \sqrt{3} A_q$, and the pre-Melvinized 5D metric and gauge field $A_q$ are solutions to the equation of motion coming from this action.

We can now apply the Melvin machine to this 10d solution.  Using $e^t$ and $e^\xi$ (or $e^\tau$ and $e^y$)  from (\ref{BHFive}) and defining $e^\chi \equiv \eta +  A_{q (m)} dx^m$ to be our 1-forms (\ref{OneForms}), we find the resulting 10d solution,
\begin{eqnarray}
\label{MelMet}
{ds^2_{10}}' &=& {ds^2_{5}}(M)'+ ds^2_{X} +  e^{2V} (\eta + A_q)^2 \,,\\
F_5' &=& F^0_{5}+B'_{2} \wedge F'_{3} \,,\\
\label{MelF} F_3' &=& f \wedge d\eta\,, \\
\label{MelB} B_2' &=& A_M \wedge (\eta + A_q)\,, \\
\Phi' &=& \Phi_0 - {1 \over 2} \log K\,,
\label{MelDil}
\end{eqnarray}
where the new five-dimensional metric takes the form
\begin{eqnarray}
{ds^2_{5}}(M)' =  {||G_{\alpha\beta}||\over K} e^{t} e^{t}+ {1\over K}G_{\a\b}\, e^{\a}e^{\b} + G_{mn}\, dx^m dx^n \,, \label{FiveMelMet}
\end{eqnarray}
with $||G_{\alpha \beta}|| \equiv G_{tt} G_{\xi\xi} - G_{t\xi}^2 =G_{\tau\tau} G_{yy} - G_{\tau y}^2$, and 
\begin{eqnarray}
K =  e^{-2V} = 1 + G_{\xi\xi} \,,
\end{eqnarray}
and where we have defined 1-forms $A_M$ and $f$ on $M$:
\begin{eqnarray}
\label{AMandf}
A_M = -e^{2V}\!  G_{\xi\a} e^{\a} \,, \quad \quad f = - {1 \over 2}  dA_{q(\xi)} \,.
\end{eqnarray}
Importantly, the vector $A_q$ does not change.  

In the original $\tau$-$y$ coordinates, this takes the equivalently simple form,
\begin{eqnarray}
{ds^2_{5}}(M)' &=&  { \beta^2 ||G_{\alpha\beta}||\over K} (e^\tau + e^y)^2+ {1\over K}G_{\a\b}\, e^{\a}e^{\b} + G_{mn}\, dx^m dx^n \,,\label{FiveMelMetty} \\
K &=&  e^{-2V} = 1 + \beta^2 (G_{\tau\tau}  + G_{yy} - 2 G_{\tau y}) \,, \\
A_M &=& - \beta e^{2V}\!  (G_{y\a} - G_{\tau \a} )e^{\a} \,, \quad \quad f = - {\beta \over 2}  d(A_{q(y)} - A_{q(\tau)}) \,,
\end{eqnarray}
where now $\alpha, \beta$ run over $\tau$ and $y$.

Notice that the Melvinized 5D metric does not involve $A_q$ or any other data, but only components of the original 5D metric tensor.  In the case where $G_{t\xi} = 0$, one can see that
\begin{eqnarray}
\label{GttUnchanged}
G'_{tt} = {1 \over K} \( G_{tt} + G_{tt} G_{\xi\xi} \) = G_{tt} \,,  \quad \quad (G_{t \xi} = 0) \,,
\end{eqnarray}
and hence only the $G'_{\xi\xi} = G_{\xi\xi}/(1 + G_{\xi\xi})$ component changes in this case.  The generalization of this observation to $G_{t\xi} \neq 0$ can be found in appendix~\ref{DimRedApp}.

The metric also determines the new vector field $A_M$, while the gauge field $A_q$ fixes $f$.  The nature of the vector $A_M$ can be understood more easily if we raise its index, with either the pre-Melvinized or post-Melvinized 5D metric.  Either way we find it points purely in the $\xi$-direction:
\begin{eqnarray}
\label{AMKilling}
G'^{\mu\nu} (A_M)_\nu = -  \delta^\mu_\xi = e^{-2V} G^{\mu\nu} (A_M)_\nu\,.
\end{eqnarray} 
In particular, this implies that $A_M$ is (minus) the $\xi$-Killing vector with respect to the Melvinized metric.

We now turn to dimensionally reducing the 10D type IIB supergravity action to five dimensions using (\ref{MelMet})-(\ref{MelDil}) 
as an ansatz for the fields; see Appendix C for further details.  Since $V = \Phi$ in our solution, the simplest dimension reduction involves setting them equal and keeping only this single scalar field.\footnote{We have set $\Phi_0 = 0$ for convenience; it can easily be restored if necessary.}  We find the (string frame) result,
\begin{eqnarray}
\label{FiveAction}
2 \kappa_5^2\, S_5 &=&  \int {\rm vol}_M e^{-  \Phi} \Bigg[ {R'}^{(5)} 
+ 16 - 4 e^{ 2\Phi}  \Bigg] 
 - {1 \over 2} \int e^\Phi F_q \wedge *_5' F_q\\&& - \int e^{-\Phi}(F_q - 2 f \wedge A_M) \wedge *_5' (F_q - 2 f  \wedge A_M) + \int A_q \wedge F_q \wedge F_q  - {1 \over 2}\int 8 e^\Phi f \wedge *_5' f  \nonumber \\ && 
 - {1 \over 2}\int \Bigg[ e^{-\Phi} (A_M \wedge F_q) \wedge *_5' (A_M \wedge F_q)
 + e^{- 3 \Phi} F_M \wedge *_5' F_M + 8 e^{ - \Phi} A_M \wedge *_5' A_M \Bigg] \,, \nonumber
 \end{eqnarray}
 with $2 \kappa_5^2 \equiv 2 \kappa_{10}^2/ {\rm vol}(S^5)$.
 Thus upon KK reduction on $Y$, the five-dimensional effective dynamics includes the 
 metric ${ds^{2}_{5}}'$; a massless vector $A_{q}$, which was present in the 10D metric before the Melvin map acted; a massive vector $A_{M}$ coming from $B_2$ of the type that is by now standard in Schr\"odinger-type backgrounds; and a scalar $\Phi$.   Upon setting $\Phi = V$ the kinetic term for the combined field vanishes, but it will reappear in Einstein frame.

 Note that $f$, the field characterizing $F_3$, is completely fixed by the vectors $A_{M}$ and $A_{q}$ and is not an independent mode. 
The field equation for $f$ involves no derivatives, consistent with what we obtain from reducing the ten-dimensional $F_3$ equation of motion.  Varying (\ref{FiveAction}) with respect to $f$, we obtain
\begin{eqnarray}
 f &=&   - {1 \over 2} e^{-2 \Phi} *_5' \Big(A_M \wedge *_5' (F_q - 2 f \wedge A_M) \Big) \,, \\
 &=& - {1 \over 2} e^{-2 \Phi} *_5 (A_M \wedge *_5 F_q) \,,
\end{eqnarray}
where in the second line we used the results (\ref{HodgeStarFq}), (\ref{HodgeXi}) and (\ref{AMNon}); evaluating the Hodge stars and using that $A_M$ is a Killing vector for $\xi$ (\ref{AMKilling}) gives us precisely the expression (\ref{AMandf}) for $f$.

To pass to five-dimensional Einstein frame, we define
\begin{eqnarray}
G^E_{\mu\nu} \equiv e^{-2 \Phi/3} G_{\mu\nu} \,.
\end{eqnarray}
This leads to the 5D Einstein frame action,
\begin{eqnarray}
\label{FiveEinsteinAction}
2 \kappa_5^2 \,S_{5,E} &=&  \int {\rm vol}^E_M \Bigg[ {R'}_E^{(5)} - {4 \over 3} \partial_\mu \Phi \partial^{\mu_E'} \Phi + 16 e^{2\Phi/3} - 4 e^{ 8\Phi/3}  \Bigg] 
 - {1 \over 2} \int e^{4\Phi/3} F_q \wedge *_E' F_q\\&& - \int e^{-2\Phi/3}(F_q - 2 f \wedge A_M) \wedge *_E' (F_q - 2 f  \wedge A_M)   \nonumber \\ && + \int A_q \wedge F_q \wedge F_q  - {1 \over 2}\int 8 e^{2\Phi} f \wedge *_E' f  \nonumber \\ && 
 - {1 \over 2}\int \Bigg[ e^{-4\Phi/3} (A_M \wedge F_q) \wedge *_E' (A_M \wedge F_q)
 + e^{- 8 \Phi/3} F_M \wedge *_{E}' F_M + 8  A_M \wedge *_E' A_M \Bigg] \,. \nonumber
 \end{eqnarray}
The limit of the action in either frame (\ref{FiveAction}) or (\ref{FiveEinsteinAction})  where $A_M = \Phi = f =0$ is simply minimal gauged supergravity (\ref{FiveActionTrunc}).

\section{Properties of Melvinized Black Holes}

In this section, we study the general properties of generic Schr\"odinger black holes which arise by Melvinization, including their thermodynamics, extremal limits and near-horizon geometries.  

It was shown in \cite{MMT,HRR,ABM} that Melvinizing the Schwarzschild-$AdS_{5}$ black hole leads to a special asymptotically Schr\"odinger black hole whose odd thermodynamic properties are a consequence of its origin.\footnote{This unusual thermodynamics, which we will recapitulate below, was shown to similarly follow from DLCQ in the dual field theory in an interesting paper by Barbon and Fuertes\cite{BarbonFuertes}; see also \cite{KovtunNickel} for a discussion of the more lenient constraints of Schr\"odinger symmetry on the thermodynamic potentials.}    As we shall see, a number of features of the charged and rotating asymptotically Schr\"odinger black holes derived via Melvinization are similarly inherited from their pre-Melvin progenitors-- in particular, their temperature, entropy, chemical potential and properties of the near-horizon geometry.

Consider a black hole metric of the class in (\ref{BHFive}), with the particular form
\begin{eqnarray}
ds^2_5(M) = G_{\tau\tau} (e^{\tau})^2 + 2 G_{y \tau} e^y e^\tau + G_{yy} (e^y)^2 + G_{rr} dr^2 + G_{ij} dx^i dx^j \,,
\end{eqnarray}
where we choose Schwarzschild-like coordinates where $G^{rr} \to 0$ at the horizon $r = r_+$.  We use $\tau$ and $y$, rather than $t$ and $\xi$, as the more natural coordinates near the horizon.  The Melvinized solution is obtained from (\ref{FiveMelMetty}), and passing to Einstein frame produces an extra factor of $K^{1/3}$:
\begin{eqnarray}
{ds^2_5(M)}' &=& {1 \over K^{2/3}} \Big( G_{\tau\tau} (e^{\tau})^2 + 2 G_{y \tau} e^y e^\tau + G_{yy} (e^y)^2 + \beta^2 (G_{\tau\tau} G_{yy} - G_{y \tau}^2) (e^\tau + e^y )^2 \Big) \\ &&+ K^{1/3} G_{rr} dr^2 + K^{1/3} G_{ij} dx^i dx^j \,, \nonumber
\end{eqnarray}
where as before
\begin{eqnarray}
K = 1 + \beta^2 (G_{\tau\tau} + G_{yy} - 2 G_{y \tau}) \,.
\end{eqnarray}
In general $K$ is finite and nonzero at the horizon; as long as this holds,  the horizon still exists at the same radius ${G'}^{rr}(r = r_+) = 0$ after the Melvin map is performed.  We will now enumerate a number of thermodynamic properties related to this horizon.

For simplicity, here we will consider the class of black holes where $e^\tau = d\tau$.  This restricts the only possible rotation of the hole to be along the $y$-direction, which encompasses our examples.  There is no reason this analysis cannot be generalized.

The divergence of $G_{rr} \to \infty$ at the horizon is in general accompanied by a corresponding zero in the determinant of $G$, so that the volume element remains finite at the coordinate singularity.  When $G_{\tau y} = 0$, it is  $G_{\tau\tau} \to 0$ that provides this zero;
when $G_{\tau y}$ is nonzero, $G_{\tau\tau}$ vanishes at the stationary limit surface, which may not coincide with the horizon, and instead the determinant of the $\tau$-$y$ metric vanishes at the horizon. To emphasize this we can write the metric in the non-coordinate form
\begin{eqnarray}
\label{NonCoordTau}
ds^2_5(M) &=& {\cal G}_{\tau\tau} d\tau^2 + {\cal G}_{yy} \Big(e^y  + {G_{\tau y} \over G_{yy}} d\tau\Big)^2 +G_{rr} dr^2+ G_{ij} dx^i dx^j \,, \\
{\cal G}_{\tau\tau} &\equiv& {G_{\tau\tau} G_{yy} - G_{\tau y}^2 \over G_{yy}} \,, \quad \quad {\cal G}_{yy} \equiv G_{yy} \,. \nonumber
\end{eqnarray}
In general at the horizon ${\cal G}_{\tau\tau} \to 0$, and we shall assume this is the case in what follows.

\subsection{Entropy}
The entropy of the black hole is simply proportional to the area of the horizon, $S = A/4G_5$, integrated at constant $\tau$ and $r = r_+$.  For the unMelvinized solution, this reads
\begin{eqnarray}
S = {1 \over 4 G_5} \int \sqrt{G_3} =  {1 \over 4 G_5} \int \sqrt{G_{yy} \,{\rm det}(G_{ij})}\,.
\end{eqnarray}
In the Melvin case we get
\begin{eqnarray}
S' &=& {1 \over 4 G_5} \int \sqrt{G_3'} =  {1 \over 4 G} \int \sqrt{K^{-2/3} (G_{yy} + \beta^2 (G_{\tau\tau} G_{yy} - G_{\tau y}^2)) \,{\rm det}( K^{1/3} G_{ij})}\,, \\
&=&  {1 \over 4 G_5} \int \sqrt{G_{yy} \,{\rm det}(G_{ij})} = S \,,
\end{eqnarray}
where we used $G_{\tau\tau} G_{yy} - G_{\tau y}^2 = 0$ at the horizon, and found the factors of $K$ to cancel.  Thus the area of the horizon, and consequently its entropy, is unchanged by the Melvin map.

\subsection{Temperature and Chemical Potential}

The temperature is most easily calculated by analytically continuing to imaginary time and verifying that the region near the horizon in the $\tau$-$r$ plane is free of conical singularities.  This is conveniently done with the metric in the form (\ref{NonCoordTau}), and we find
\begin{eqnarray}
T = \lim_{r \to r_+} {1 \over 2 \pi \sqrt{G_{rr}}} {d \over dr} \sqrt{{\cal G}_{\tau\tau}} \,.
\end{eqnarray}
Under the Melvin map, the relevant quantities transform like (in Einstein frame),
\begin{eqnarray}
{\cal G}'_{\tau\tau} = K^{1/3} {{\cal G}_{\tau\tau} \over 1 + \beta^2 {\cal G}_{\tau\tau}} \,, \quad \quad
G'_{rr} = K^{1/3} G_{rr} \,,
\end{eqnarray}
and using the vanishing ${\cal G}_{\tau\tau}(r \to r_+) \to 0$ at the horizon,
\begin{eqnarray}
 {d \over dr} \sqrt{{\cal G}'_{\tau\tau}} = K^{1/3} \(  {d \over dr} \sqrt{{\cal G}_{\tau\tau}}\) \left( 1 + {\cal O}({\cal G}_{\tau\tau}) \right) \,, 
\end{eqnarray}
where we assumed the regularity of $dK/dr$ at the horizon, implying
\begin{eqnarray}
T' = \lim_{r \to r_+} {1 \over 2 \pi \sqrt{G'_{rr}}} {d \over dr} \sqrt{{\cal G}'_{\tau\tau}}  = T \,,
\end{eqnarray}
and thus the temperature is unchanged by the Melvin map.

This result assumes the temperature is defined with respect to the same Killing generator of the event horizon both before and after the Melvin transformation.  One does have to take into account the change from using $i \partial_\tau$ as the asymptotic time coordinate, as is suitable for an AdS black hole, to using $i \partial_t$ as the asymptotic time coordinate as suits the Schr\"odinger cases.

In our class of solutions, the generator of the Killing horizons can take the form
\begin{eqnarray}
\chi &=& \partial_\tau + \Omega_H \partial_y \\
&=& \beta (1 + \Omega_H) \partial_t - {1 \over 2 \beta} (1 - \Omega_H) \partial_\xi \,,
\end{eqnarray}
where $\Omega_H$ parameterizes rotation in the $y$-direction; $\chi$ has unit coefficient with respect to the $\tau$-coordinate.  To switch to a generator suited to the $t$-coordinate, we define
\begin{eqnarray}
\label{KillingHor}
\chi_t \equiv{1 \over \beta (1 + \Omega_H)} \chi = \partial_t - {1 \over 2 \beta^2} {1 - \Omega_H \over 1 + \Omega_H} \partial_\xi \,.
\end{eqnarray}
Since the temperature may be defined as $T \equiv \kappa/(2 \pi)$ where the surface gravity $\kappa$ is
\begin{eqnarray}
\label{SurfaceGrav}
\kappa^2 \equiv - {1 \over 2} (\nabla_\mu \chi_\nu) (\nabla^\mu \chi^\nu) \,,
\end{eqnarray}
we see that the shift in coordinates to the Schr\"odinger time $t$ produces a rescaling of the temperature,
\begin{eqnarray}
T_t = {T_\tau \over \beta (1 + \Omega_H)} \,.
\end{eqnarray}
We can also read the chemical potential off the expression (\ref{KillingHor}).  Since $i \partial_t$ corresponds to the Hamiltonian $H$ while $i \partial_\xi$ is the number operator $N$, we obtain the chemical potential from the identification $\chi_t \sim H + \mu N$ for the grand canonical ensemble,
\begin{eqnarray}
\mu = {1 \over 2 \beta^2} {\Omega_H-1 \over  \Omega_H+1} \,.
\end{eqnarray}

\subsection{Near-horizon limit}

One of our main goals in studying charged and rotating asymptotically Schr\"odinger black holes is to find extremal examples with scale-invariant near-horizon geometries, or throats.  Quite apart from just being unusually simple, such extremal black holes have become important tools in the application of gauge-gravity duality in a number of contexts, most notably in the holographic description of (non-)fermi liquids \cite{LMV}.\footnote{The reason throats are important is easy to see.  When the near-horizon region has a scaling invariance (for example, the near-horizon $AdS_{2}$ symmetry of the extremal RN-$AdS_{4}$ black holes), taking a scaling limit allows one to consistently decouple the modes near the horizon from the modes near the boundary.  Since the horizon and boundary are holographically related to IR and UV physics, this means that at least some aspects of UV and IR physics may be studied independently of each other, a powerful statement of universality.  This view is elegantly presented in \cite{FILM} and by Liu in a talk at the KITP on July 7, 2009.  AA thanks H.~Liu and J.~McGreevy for explaining their work, and D.~Marolf and A.~Sinha for discussion on the meaning of the near-horizon geometry of extremal black holes.}

A number of extremal black holes are known to have near-horizon AdS$_2$ regions.  We now show that if this is the case for the pre-Melvin extremal metric, the Melvinized result also has an AdS$_2$ region, albeit with a different radius of curvature.
Recall that the $AdS$ portion of the near-horizon limit of the original $AdS_{5}$ black hole came from the ${\cal G}_{\tau\tau} d\tau^2 + G_{rr} dr^2 $ portion of the metric. Under the Melvin map above, this sector becomes,
\begin{eqnarray}
{\cal G}_{\tau\tau} d\tau^2 + G_{rr} dr^2 \quad \to \quad
K^{1/3} \(  {\cal G}_{\tau\tau}( 1 + {\cal O}({\cal G}_{\tau\tau}))  d\tau^2 + G_{rr} dr^2 \) \,.
\end{eqnarray}
The near-horizon geometry is thus simply rescaled by a factor of $K^{1/3}$ common to both terms.  In the case of an extremal black hole with a near-horizon $AdS_2$ region, the Melvinized Schr\"odinger version will have an analogous $AdS_2$ region, with the $AdS$ radius rescaled by the factor $K^{1/6}$.  

Due to the different scalings of the $y$ direction from the $x$ and $z$ directions under Melvin, the remaining three-dimensional space is in general squashed by a factor of $K$.  In the planar limit, this is a completely trivial rescaling of the flat spatial section.   When the spatial section is $S^{3}$, however, the net result is a squashing of the $S^{3}$.

One interesting consequence of this result is that the radius of curvature of the near-horizon region is now independent of that of the asymptotic region -- explicitly, $K$ depends on $\b^{2}$, which can be tuned independently.  That the radius of curvature of the extremal throat depends explicitly on the density of non-relativistic excitations in the boundary theory is an interesting result, and may play in important role in understanding holography in this spacetime.

\section{Examples: Schr\"odinger black holes with charge}

In this section, we apply our Melvin map to a pair of five-dimensional AdS black holes, and explicitly write down the resulting backgrounds.  We also discuss thermodynamic properties, and find the results consistent with the previous section where we showed they are unchanged by Melvin, modulo the coordinate transformations needed for asymptotically Schr\"odinger solutions.
First we consider a charged AdS black hole in the Poincar\'e patch; this will give us the simplest charged Schr\"odinger black hole.  Then we examine the more complicated case of a charged black hole in global coordinates, rotating along the Hopf fiber of the global $S^3$; this includes as a limiting case the chargeless rotating black hole.

\subsection{Charged Schr\"odinger black hole}

As a first application of our results, we can construct the first charged black hole with asymptotic Schr\"odinger symmetries, which we call RN-Sch$_5$ for Reissner-Nordstrom-Schr\"odinger in five dimensions.

Our pre-Melvinized solution is a five-dimensional RN-AdS$_5$ black hole.  This solution is associated to a ten-dimensional solution of D3-branes rotating around the Hopf direction $\chi$ of the $S^5$ \cite{Chamblin:1999tk, Cvetic:1999xp}.    The five-dimensional metric and gauge field are
\begin{equation}\label{eq:qBH10}
ds^2 =
\frac{1}{r^2}\left[-\frac{f}{H^2}d\tau^2+H\left(\frac{dr^2}{f}+d\vec{x}^2\right)\right]
\,, \quad \quad A_q = {q r^2 \over H r_H^2} d\tau\,,
\end{equation} 
where $q$ is the charge, and
\begin{eqnarray}
 f \equiv H^3-\frac{r^4}{r_H^4} \,,  \quad \quad H \equiv 1+ q^2 r^2 \,. 
 \end{eqnarray}
Note that $r_H$ is the location of the horizon in the {\it un}charged geometry; it is related to the mass parameter $m$ by $r_H^4 =1/2m$.  In what follows
we use $r_+$ to denote the true horizon radius.
 This background is a solution to the action (\ref{FiveActionTrunc}) for five-dimensional minimal gauged supergravity, and lifts to a ten-dimensional solution of type IIB supergravity described by (\ref{BHTen}), (\ref{F5Ten}) with $X = \IP^2$ and $\eta = \eta^{(5)}$ the Hopf fiber generating the map $S^1 \to S^5 \to \IP^2$; see appendix~\ref{CoordApp} for more details.

We now turn the Melvinization crank as outlined in the section~\ref{MelvinSec}, with $e^\tau = d\tau$ and $e^y = dy$; the directions $x$, $y$, $z$ are all equivalent and so the choice of Melvin direction is arbitrary.
We arrive at the five dimensional charged Schr\"odinger solution in Einstein frame:
\begin{equation}\label{eq:qBH5}
ds^2 = \frac{K^{1/3}}{r^2}\left[-{f \over H^2 K} d\tau^2-\frac{f\beta^2
    }{r^2 H K}\left(d\tau+dy\right)^2+{H \over K} dy^2 + H \left( \frac{dr^2}{f}+dx^2 + dz^2 \right) \right] \,,
\end{equation}
with
\begin{eqnarray}
K = 
1+\frac{\beta^2 r^2}{H^2 r_H^4}\,.
\end{eqnarray}
The overall factor of $K^{1/3}$ is absent in string frame.
For the special
case $q=0$, (\ref{eq:qBH5}) is precisely the uncharged Schr\"odinger geometry
previously studied in \cite{HRR,MMT,ABM}.

\subsubsection*{Entropy}
The entropy $S$ associated with (\ref{eq:qBH5}) is easily computed via
the Beckenstein-Hawking result $S = A/4 G_5$ where $A$ is the
area of the horizon, and $G_5$ is the gravitational constant in five
dimensions. Performing this calulation, we find for the entropy density,
\begin{equation}\label{eq:sqBH}
s \equiv \frac{S}{V_{x,y,z}}=\frac{1}{4G_5\, r_H^2r_+} = \frac{H(r_+)^{3/2}}{4 G_5 r_+^3}
\end{equation}
where $V_{x,y,z}$ is the (infinite) horizon volume. Again, in the
limit $q\to 0$ ($r_+ \to r_H$), this result agrees with those obtained in \cite{HRR,MMT,ABM}.  Furthermore, it is manifestly independent of $\beta$, demonstrating that it was not changed by the Melvin procedure, consistent with our general argument in the previous section.

\subsubsection*{Temperature and Chemical Potential}
The corresponding Hawking temperature is readily calculated from the relation $T=
\kappa/2\pi$ where $\kappa$ is the surface gravity of the
horizon (\ref{SurfaceGrav}). As discussed in the previous section, to obtain the temperature associated with the asymptotic Schr\"odinger Hamiltonian $H = i \partial_t$ we should take the generator the horizon to be $\chi_t  = \partial_\tau/\beta$; we have $\Omega_H = 0$ in this nonrotating case.
We then
find
\begin{equation}\label{eq:tqBH}
T = \frac{\Big |f'(r_+)\Big |}{4\pi\beta}\left(\frac{r_H}{r_+}\right)^2 = \frac{\sqrt{H(r_+)}}{2\pi\beta\, r_+}\Big | H(r_+) -3 \Big| \,.
\end{equation}
Notice that the temperature vanishes when $H(r_+) = 3$, which implies the geometry becomes extremal when
\begin{equation}\label{eq:extremal}
q = \frac{\sqrt{2}}{r_+} \,.
\end{equation}
We will use this fact in the following section.  The temperature corresponding to the asymptotic $\tau$-coordinate is $\beta T$, which is manifestly $\beta$-independent and thus unchanged by Melvin.

Because $\chi_t$ takes the form
\begin{equation}\label{eq:kg}
\chi_t = \frac{1}{\beta}\partial_\tau=\partial_t-\frac{1}{2\beta^2}\partial_\xi \,,
\end{equation}
we can read off the chemical potential for for $\xi$-translations as
\begin{equation}\label{eq:chemPot}
\mu = -\frac{1}{2\beta^2} \,,
\end{equation}
consistent with the previous section.

\subsubsection*{On Shell Action}
A straightforward, if tedious, application of familiar techniques from holographic renormalization allows us to construct a renormalized on-shell action for this solution.  Dispensing with the technicalities, the result is,
\begin{eqnarray}
\label{finally}
S_{5}=-{1\over 16 \k_{5}}\int dx^4 {H[r_+]^3(1-2q^2\beta^2)\over r_+^4} \,.
\end{eqnarray} 
Continuing to periodically identified euclidean time then gives the on-shell Euclidean action,
\begin{eqnarray}
S_{E}=-{\pi\b V\over 8 \k_{5}}{H[r_+]^{5/2}(1-2q^2\beta^2)\over  r_+^3 \, | H(r_+) -3 | } \,.
\end{eqnarray} 
which we would like to identify as the action of a saddle-point approximation to the full grand-canonical partition function.

\subsubsection*{The Extremal, Near-Horizon Limit}
Near the horizon of an extremal Reissner-Nordstrom black hole, the geometry becomes a direct product space with one factor AdS$_2$.  We expect an analogous limit for 
 the extremal version of (\ref{eq:qBH5}).  To obtain this, we define the deviation from the horizon $\zeta$,
 \begin{eqnarray}
 \zeta \equiv {r \over r_+} - 1\,,
 \end{eqnarray}
 and expand the metric (\ref{eq:qBH5}) in the extremal limit (\ref{eq:extremal}) in powers of $\zeta$.  The horizon in the extremal case is located at
 \begin{equation}
r_+ = 27^{1/4} \,  r_H= \left( \frac{27}{2m} \right)^{1/4} \,.
\end{equation}
 Shifting $\tau$ to absorb the cross term with $y$, rescaling the coordinates and inverting $\zeta = 1/\rho$, we find the near-horizon result
\begin{equation}
ds^2 = \frac{L^2}{\rho^2}\left( d\tau^2 + d\rho^2\right) + dx^2 + dy^2 + dz^2 \,.
\end{equation}
 Lo and behold, the limiting form  is AdS$_2 \times \mathbb{R}^3$, with AdS radius
\begin{equation}
L^2 = {K(r_+)^{1/3} \over  12} = \frac{\left( 1+ 3\beta^2/r_+^2\right)^{1/3}}{12} \,,
\end{equation}
which as anticipated, depends on $\beta$.

\subsection{Charged, Hopf-rotating Schr\"odinger black hole}

We now turn to a more complicated example, a charged AdS black hole that is also rotating.  In five dimensions there are two independent rotation parameters, in the literature usually termed $a$ and $b$; we focus on the case where $b = a$, corresponding to rotation along the Hopf fiber of the $S^3$ of global AdS.  The uncharged version of this solution was written in \cite{HHT}, while the charged version, which is a solution to minimal gauged supergravity (\ref{FiveActionTrunc}) was formulated in \cite{Cvetic}.  The solution reads in our conventions
\bea
ds^{2}&=&
-{q+\s^{2}(1+\r^{2}/L^2)\over\s^{2}\Delta}d\tau^{2}+{f+q\s^{2}\Delta\over\s^{4}\Delta^{2}}\(d\tau-a\,\eta^{(3)}\)^{2} \\ \non
&&+ {\s^{2}\over \Delta_{\r}}d\r^{2} + {\s^{2}\over \Delta}\[ds^{2}_{\IP^{1}}+\(1+{q\,a^{2}\over\s^{4}}\)(\eta^{(3)})^{2}\] \,,\\
A_q&=&{q\over \s^{2}\Delta}\(d\tau-a\, \eta^{(3)}\) \,,
\eea
where $\eta^{(3)}$ is the vertical 1-form along the Hopf fibre of the $S^{3}$, as described in appendix~\ref{CoordApp}, and 
\begin{eqnarray}
\Delta&=&1-a^2/L^2 \,,
~~~~~~
\s^{2}=\r^{2}+a^2 \,,
~~~~~~
f=2m\,\s^{2}+2q\,a^2\s^{2}/L^2-q^{2} \,, \\
\Delta_{\r}&=&{q^{2}+2q a^2+\s^{4}(1+\r^{2}/L^2) - 2m\r^{2} \over \r^{2}} \,. \nonumber
\end{eqnarray}
In the limit $a \to 0$ this solution becomes the RN-AdS$_5$ black hole in global coordinates, while for $q \to 0$ it reduces to the uncharged AdS-Kerr black hole rotating along the Hopf fiber.

This solution is readily Melvinized.  We take $e^\tau =d\tau$, $e^\psi = 2 \eta^{(3)} = d\psi + \cos \theta d\phi$, where $\psi$ takes the place of $y$ as the T-duality direction; we are thus choosing the Hopf direction for the $y$-isometry.  The resulting Melvinized Kerr-Newman-Sch$_{5}$ metric is then
\bea
ds^{2}&=&
-{q+\s^{2}(1+\r^{2}/L^2)\over K \s^{2}\Delta}d\tau^{2}+{f+q\s^{2}\Delta\over K \s^{4}\Delta^{2}}\(d\tau-a\,\eta^{(3)}\)^{2}+ {\s^{2}\over K \Delta} \(1+{q\,a^{2}\over\s^{4}}\)(\eta^{(3)})^{2} \\ \non
&& -\frac{\beta^2 \rho^2\, \Delta_\rho}{4 \Delta^2 \sigma^2 K} (d\tau + 2 \eta^{(3)})^2+ {\s^{2}\over \Delta_{\r}}d\r^{2} + {\s^{2}\over \Delta} ds^{2}_{\IP^{1}}\,,
\eea
where
\begin{eqnarray}
K = 1 + \beta^2 \( {f (2+a)^2 \over4  \Delta^2 \sigma^4} + { a (2 + a) q \over 2  \Delta \sigma^2} + {\sigma^2 - 4 (1 + \rho^2/L^2) \over 4 \Delta}      \)\,.
\end{eqnarray}

\subsubsection*{Entropy, Temperature and Chemical Potential}
The entropy density associated with this geometry is found to be
\begin{equation}
\label{Entropyaq}
s = \frac{\sigma_+^4 + a^2 q}{4 G_5\, \Delta^2\, \rho_+}
\end{equation}
which, as expected, is independent of $\beta$ and hence identical to that obtained from the 5D {unMelvinized metric. As a result, it agrees with \cite{Cvetic}.

To determine the temperature and chemical potential, it is necessary to compute the null generator of the horizon for this geometry. Because the metric is stationary but not static, the generator will be of the form $\chi \propto \partial_\tau + \Omega_H \partial_\psi$ where $\Omega_H$ is the angular velocity of a test particle at location $\rho_+$.
By considering a photon emitted in the $\psi$ direction at fixed $\rho$, $\theta$, and $\phi$, it is easy to show that,
\begin{equation}\label{eq:omega}
\Omega_H = \frac{2a\left( \sigma_+^2\left(1+\rho_+^2\right)+q \right)}{\sigma_+^4+a^2 q} \,.
\end{equation}
This too is unchanged by Melvinization. As is by now familiar, to use our null generator for asymptotically Schr\"odinger thermodynamics, one must normalize it such that the coefficient of $\partial_t$ is one. As in (\ref{KillingHor}), this amounts to a scaling by $1/\beta(1+\Omega_H)$:
\begin{equation}
\chi_t = \partial_t + \frac{1}{2\beta^2}\left(\frac{\Omega_H - 1}{\Omega_H +1}\right)\partial_\xi\,,
\end{equation}
from which we easily read off the chemical potential
\begin{equation}
\label{ChemPotaq}
\mu =  \frac{1}{2\beta^2}\left(\frac{\Omega_H-1 }{\Omega_H +1}\right) \,.
\end{equation}
We note that the rotation of the original AdS black hole has been converted into a rescaling of the chemical potential for the corresponding Schr\"odinger geometry.  In particular, there is no spatial rotation associated to the $(2+1)$-dimensional field theory dual spacetime.  Such a rotation could be obtained by considering the Melvin map on an AdS black hole with two independent angular momenta.

The temperature must also be identical to that of the un-Melvinized $5d$ geometry \cite{Cvetic}, up to a rescaling introduced by the properly normalized null generator, and we find
\begin{equation}\label{eq:Taq}
T = \frac{1}{2\pi \beta}\frac{\rho_+^4\left( 1+2\sigma_+^2\right)-(a^2+q)^2}{\rho_+\left(1+\Omega_H\right) (\sigma_+^4 +a^2 q)} \,.
\end{equation}

\subsubsection*{The Extremal, Near Horizon Limit}
From (\ref{eq:Taq}), it is easy to see that the geometry becomes extremal when 
\begin{equation}\label{eq:qexAQ}
q = -a^2 \pm \rho_+^2\sqrt{1+2\sigma_+^2} \,.
\end{equation}
Using this fact, we can again study the near horizon limit of the extremal solution, whose horizon is located at 
\begin{equation}
\rho_+ = \frac{1}{\sqrt{3}}\left[ \sqrt{1+6 m +a^2(a^2-2)}-1-2 a^2 \right]^{1/2} \,.
\end{equation}
As before, we scale the radial coordinate like $\rho/\rho_+ = 1+\zeta$ and expand about $\zeta =0$. Although the details are tedious, we again find all geometries satisfying (\ref{eq:qexAQ})
 decompose into a direct product space with one factor AdS$_2$. Computationally, it is easiest to work with the metric written in the  form (\ref{NonCoordTau}), in which case $\mathcal{G}_{\tau\tau}\propto r^2$ and $g_{\rho\rho}\propto r^{-2}$ near the horizon, with all other components $\mathcal{O}(r^0)$.

\section{Conclusion}

In this paper we have obtained a general expression for the action of the null Melvin twist, constructed a series of charged and rotating asymptotically Schr\"odinger black hole solutions of type IIB supergravity, found a 5d truncation of the 10d IIB theory adapted to these solutions, and examined some of the salient features of their geometry and thermodynamics.  Along the way we discovered that these spacetimes inherit extremal limits from their parent AdS spaces, but that some of their features -- like the radii of curvature of the near-horizon regions -- are not inherited, leading to a potentially interesting class of new solutions.  The fact that these systems are explicitly embedded in IIB string theory allows us to identify the boundary CFT as the $_{\b}$DLCQ of the pre-Melvin boundary theory.  Such a theory is by construction a non-relativistic conformal field theory.

Whatever this boundary system is, however, it is decidedly not a good description of fermions at unitarity.  First, the thermodynamics of this system is wrong.  For example, while the pressure is positive, it scales with a negative power of the chemical potential, ${\cal P} \sim {T^{4} / \mu^{2}}$; positivity of the number density $N={\p{\cal P} / \p\m}$ then requires that the chemical potential is negative.  Notably, this scaling is not a consequence of NR conformal symmetry, as the conformal algebra requires only that ${\cal P} \sim T^{d+2\over2}g({\mu / T})$, as elegantly explained in \cite{KovtunNickel}.  Rather, this odd thermodynamics is a direct consequence of the identification of particle number with a DLCQ momentum, $N=i\p_{\xi}$, as powerfully argued in \cite{BarbonFuertes}, who found that the curious scalings follow from the summation over the infinite tower of KK modes.  

Secondly, and importantly, while the system certainly has a finite density of non-relativistic excitations, it is not in a superfluid phase, since the $U(1)$ conjugate to the particle number -- here, $\xi$-translation invariance -- is manifestly {\em un}-broken.  It is certainly possible that this geometry is secretly a subleading saddle, with a $\xi$-momentum violating solution dominating entropically; unfortunately, no such solution is presently known.  

Eventually, one would like to find solutions that dispense with the $\xi$-translation symmetry entirely, for example by breaking it explicitly even in the asymptotic regime, or by KK-reducing on a finite-density geometry in which the $\xi$-circle is spacelike and lifting the KK modes by moving out along a Coulomb branch, as was done for example in the DLCQ of M-theory. This is an important direction for future work.\footnote{We thank J.~McGreevy and D.~Son for illuminating discussions on this topic.}

All that said, even if these systems are not fermions at unitarity, they are NRCFTs and deserve study in their own right.  The solutions presented and explored in this paper open the door to a number of physically interesting questions, such as the study of superfluids and fermi surfaces in field theories which are microscopically non-relativistic.  We will return to these questions in the future.

\vspace{0.5cm}
{\bf Note Added:} While this work was being completed, we learned of work by Imeroni and Sinha \cite{Sinha} which partially overlaps with ours.

\section*{Acknowledgments}

We would like to thank
S.~de Alwis, 
G.~Horowitz, 
S.~Kachru, 
D.~Marolf,
A.~Sinha,
M.~Shultz,  
D.~Son,
and J.~Wang
for helpful discussion.  
AA and OD thank the Aspen Center for Physics and the Kavli Institute for Theoretical Physics for hospitality during the late stages of this work.  
The work of AA is supported in part by funds provided by the U.S. Department of Energy (D.O.E.) under cooperative research agreement DE-FG0205ER41360.
The work of CMB, OD and CR is supported in part by funds provided by the U.S. Department of Energy (D.O.E.) under grant DE-FG02-91-ER-40672.  

\appendix


\section{Coordinates on $S^3$ and $S^5$}
\label{CoordApp}

In our examples, we employ a Hopf parameterization of the coordinates on the compact $S^5$ and, in the cases built on global $AdS$, the $S^3$.  The metrics may be elegantly expressed in terms of the left-invariant one-forms of SU(2),
\bea 
\s_{1} = {1 \over 2}\! \(\cos\psi d\t +\sin\psi \sin\t d\phi \) \,, \;\;
\s_{2} = \half\! \(\sin\psi d\t   -\cos\psi \sin\t d\phi \) \,, \;\;
\s_{3} = \half \!\(d\psi +\cos\t d\phi \) .
\eea
The hopf-fibration $S^{1}\to S^{3} \to \IP^{1}$ enjoys the round metric,
\bea
\label{Hopf3}
ds^{2}_{S^{3}} = \s^{2}_{1} +\s^{2}_{2}  +\s^{2}_{3}  =  ds^{2}_{\IP^{1}} + (\eta^{(3)})^2    \,,
\eea
where the metric on $\IP^1$ is
\bea
ds^{2}_{\IP^{1}} = \s^{2}_{1} +\s^{2}_{2}  
={1\over 4}\(d\t^{2}+\sin^{2}\t d\phi^{2}\)\,,  
\eea
and the Hopf fiber is 
\bea
\eta^{(3)} \equiv \sigma_3  =  \half \!\(d\psi +\cos\t d\phi \)\,,
\eea
with Hopf coordinate $\psi$; thus the metric may be written
\begin{eqnarray}
ds^{2}_{S^{3}}  = {1\over4}\[  d\t^{2} +\sin^{2} \t d\phi^{2}  +\(d\psi+\cos \t d\phi\)^{2}\] \,.
\end{eqnarray}
Meanwhile the Hopf-fibration $S^1 \to S^5 \to \IP^2$ has the metric
\begin{eqnarray}
ds^2_{S^{5}} = ds^{2}_{\IP^2} +(\eta^{(5)} )^2 \,,
\end{eqnarray}
where the metric on $\IP^2$ is
\begin{eqnarray}
ds^2_{\IP^{2}} = d\mu^2 +\sin\!^2\mu \(\s_1^2+\s _2^2+\cos^2\!\mu \, \s _3^2\) \,,
\end{eqnarray}
and the Hopf fiber $\eta^{(5)}$ is
\begin{eqnarray}
\eta^{(5)} \equiv d\chi + {\cal A} \equiv d\chi + \sin^2\!\mu\,\sigma^3 \,,
\end{eqnarray}
where the Hopf coordinate is $\chi$.

In the literature on five-dimensional black holes in global $AdS$, the $S^3$ is typically written in Boyer-Lindquist coordinates $\theta_B$, $\psi_B$, $\phi_B$ rather than the Hopf coordinates $\theta$, $\psi$, $\phi$ given above.  These are related by
\begin{eqnarray}
\t = 2 \t_{B}~~~~~~\psi = \psi_{B} + \phi_{B}~~~~~~\phi = \psi_{B} - \phi_{B}\,,
\end{eqnarray}
which leads to the BL metric on $S^{3}$,
$$
ds^{2}_{S^{3}} = d\t_{B}^{2} + \sin^{2}\t_{B} d\phi_{B}^{2}+ \cos^{2}\t_{B} d\psi_{B}^{2} \,.
$$

\section{T-duality conventions}
\label{TDualApp}

We can express the T-duality rules as follows.  Let $y$ be the isometry direction along which T-duality is taken and $x^\alpha$ the remaining coordinates.  Any 10D string metric, B-field and RR fields with a $y$ isometry can be written as
\begin{eqnarray}
\label{Metric}
ds^2 &=& g_{yy} (dy + g_{(y)})^2 + {\cal G}_{\alpha \beta}\, dx^\alpha dx^\beta \,, \\
B_2 &=& \left(dy + {1 \over 2} g_{(y)} \right) \wedge B_{(y)} + {1 \over 2} {\cal B}_{\alpha \beta} \,dx^\alpha \wedge dx^\beta \,, \label{Bfield} \\
F_p &=& (dy + g_{(y)}) \wedge F_{p(y)} + F_{p(\not{y})} \,,
\label{RRfield}
\end{eqnarray}
where the one-forms $g_{(y)}$ and $B_{(y)}$ capture the off-diagonal terms between $y$ and the other directions:
\begin{eqnarray}
g_{(y)} \equiv {g_{y \alpha} \over g_{yy}} dx^\alpha \,, \quad \quad B_{(y)} \equiv B_{y \alpha} dx^\alpha \,,
\end{eqnarray}
and $F_{p(y)}$ and $F_{p(\not{y})}$ are $(p-1)$- and $p$-forms polarized along the $x^\alpha$ directions, respectively.

T-duality along the $y$-direction gives a result that may also be written in the form of (\ref{Metric})-(\ref{RRfield}),
\begin{eqnarray}
{ds^2}' &=& g'_{yy} (dy + g'_{(y)})^2 + {\cal G}_{\alpha \beta}\, dx^\alpha dx^\beta \,, \\
B_2' &=& \left(dy + {1 \over 2} g'_{(y)} \right) \wedge B'_{(y)} + {1 \over 2} {\cal B}_{\alpha \beta} \,dx^\alpha \wedge dx^\beta \,,  \\
F'_p &=& (dy + g'_{(y)}) \wedge F'_{p(y)} + F'_{p(\not{y})} \,,
\end{eqnarray}
 with
\begin{eqnarray}
g'_{yy} &\equiv& {1 \over g_{yy}} \,, \quad \quad g'_{(y)} \equiv - B_{(y)} \,, \quad \quad B'_{(y)}\equiv - g_{(y)} \,,\quad \quad e^{2\Phi'} = {e^{2\Phi} \over g_{yy}} \,,\\
F'_{p(y)} &=& F_{(p-1) (\not{y})} \,, \quad \quad F'_{p (\not{y})} = F_{(p+1)(y)} \,.
\end{eqnarray}
Note that ${\cal G}_{\alpha \beta}$ and ${\cal B}_{\alpha \beta}$ are invariant.

\section{Dimensional reduction}
\label{DimRedApp}

\subsection{Hodge dual conventions}

We use conventions for the Hodge dual where, acting on a noncoordinate basis $\hat\theta^a$,
\begin{eqnarray}
* (\hat\theta^{a_1} \wedge \ldots \wedge \hat\theta^{a_p}) = {1 \over (D-p)!} \epsilon^{a_1 \ldots a_p}_{\;\;\;\;\;\;\;\;\;\;\; a_{p+1} \ldots a_D} \hat\theta^{a_{p+1}}\wedge \ldots \wedge \hat\theta^{a_D} \,,
\end{eqnarray}
which implies
\begin{eqnarray}
F_p \wedge * F_p ={1 \over p!} F_{a_1 a_2 \ldots a_p} F^{a_1 a_2 \ldots a_p}\,  {\rm vol}  \,,
\end{eqnarray}
with ${\rm vol}$ the $D$-dimensional volume form.

For a 10D metric of the Kaluza-Klein form (\ref{BHTen}), we can split the 10D Hodge star into Hodge stars acting on the 5D $ds_5^2(M)$,  the 4D $ds^2(X)$, and the 1D $e^{2V}  (\eta^{(5)} + A_q)^2$ parts.  Conventionally ordering forms as 5D, then 1D, then 4D parts, we find
\begin{eqnarray}
*_{10} = (-1)^\delta *_5 *_1 *_4 \,,
\end{eqnarray}
where $\delta = 1$ if the  form being acted on has (even, odd, even) or (odd, even, odd) numbers of indices in the (5D, 1D, 4D) parts, and $\delta =0$ otherwise, {\em i.e.} $\delta = n_5 n_4 + n_1 (1 + n_5 + n_4)$.

We then have the useful expressions
\begin{eqnarray}
*_5 1\equiv {\rm vol}_M \,, \quad \quad *_1 1 = e^V (\eta^{(5)} + A_q) \,, \quad \quad *_4 1 = {1 \over 8} d{\cal A} \wedge d{\cal A} \,,  \quad \quad *_4 d{\cal A} = d {\cal A} \,,
\end{eqnarray}
where $J \equiv d {\cal A}/2$ is the K\"ahler form on $X$, implying 
\begin{eqnarray}
\label{10Vol}
*_{10} 1 =  e^V {\rm vol}_M \wedge \eta^{(5)}\wedge  {1 \over 8} d{\cal A} \wedge d{\cal A} \,.
\end{eqnarray}
The volume form on $Y$ is then $(*_1 1) (*_4 1)$ with $V = A_q = 0$:
\begin{eqnarray}
{\rm vol}(Y) = \int \eta^{(5)}   \wedge{1 \over 8} d{\cal A} \wedge d{\cal A} \,.
\end{eqnarray}

\subsection{Non-coordinate basis}
\label{NonCoordSec}

We would like to determine the effective five-dimensional action for which the fields of Section (2.3) provide solutions.  To do so, it is useful to write the pre- and post-Melvinized 5D metrics in a non-coordinate basis, in which the map simplifies even further.  Reexpress the original metric (\ref{BHFive}) as the $t$-$\xi$ version of (\ref{NonCoordTau}),
\begin{eqnarray}
\label{NonCoord}
ds^2_5(M) &=& {\cal G}_{tt} e^t e^t + {\cal G}_{\xi\xi} \Big(e^\xi  + {G_{t \xi} \over G_{\xi\xi}} e^t\Big)^2 + G_{mn} dx^m dx^n \,, \\
{\cal G}_{tt} &\equiv& {G_{tt} G_{\xi\xi} - G_{t \xi}^2 \over G_{\xi\xi}} \,, \quad \quad {\cal G}_{\xi\xi} \equiv G_{\xi\xi} \,,
\end{eqnarray}
so we can define non-coordinate basis 1-forms,
\begin{eqnarray}
\hat\theta^t \equiv \sqrt{{\cal G}_{tt}} e^t \,, \quad \quad
\hat\theta^\xi \equiv \sqrt{{\cal G}_{\xi\xi}} \Big(e^\xi  + {G_{t \xi} \over G_{\xi\xi}} e^t\Big) \,.
\end{eqnarray}
One may then show that ${\cal G}_{tt}$ is unchanged by Melvinization due to the cancelation of $1/K$ and the new term, generalizing (\ref{GttUnchanged}), so that $\hat\theta^t$ is fixed.  Meanwhile $\hat\theta^\xi$ changes only by an overall factor of $K^{-1/2}$.
 Thus the entire Melvin map boils down to the transformation
\begin{eqnarray}
\hat\theta^\xi \to (\hat\theta^\xi)' \equiv {1 \over K^{1/2}} \hat\theta^\xi = {1 \over \sqrt{1 + {\cal G}_{\xi\xi}}} \hat\theta^\xi \,. 
\end{eqnarray}
One can also see that $A_M$ is a 1-form proportional to $(\hat\theta^\xi)'$:
\begin{eqnarray}
\label{AMNon}
A_M = - \sqrt{ G'_{\xi\xi}}(\hat\theta^\xi)' \,.
\end{eqnarray}

This presentation allows us to understand the relationship between the 5D Hodge duals $*_5$ and $*_5'$.  When acting on a form that  contains $\hat\theta^\xi$, they differ by $e^V$:
\begin{eqnarray}
\label{HodgeXi}
e^V *_5' (\ldots \wedge \hat\theta^\xi \wedge \ldots )= *_5 (\ldots \wedge \hat\theta^\xi \wedge \ldots )\,,
\end{eqnarray}
while the converse is true when acting on a form that does not contain $\hat\theta^\xi$:
\begin{eqnarray}
e^{-V} *_5' (\ldots \wedge \not{\hat\theta^\xi} \wedge \ldots )= *_5  (\ldots \wedge \not{\hat\theta^\xi} \wedge \ldots )\,.
\end{eqnarray}
These relations are useful in understanding the presentation of the tensor $F_5^0$ (\ref{F5Ten}) after Melvinization.  This object is not changed, but it contains the pre-Melvin Hodge star $*_5$, and it is useful   for the dimensional reduction we are about to carry out to write it in terms of the post-Melvin metric.  We can easily see
\begin{eqnarray}
{\rm vol}_M \equiv *_5 (1) = e^{-V} *_5' (1) = e^{-V} {\rm vol}_{M'} \,,
\end{eqnarray}
since the Hodge stars act on a 0-form.  The $*_5 F_q$ term is not so simple, since in general $F_q$ contains both terms with and without $\hat\theta^\xi$.  However, we can use the fact that while $F_5^0$ is self-dual with respect to the unMelvinized 10D metric, only $F_5 \equiv F_5^0 + B_2' \wedge F_3'$ is self-dual with respect to the Melvin metric; one can think of the additional $B_2' \wedge F_3'$ term as being what is needed to ensure the tensor is still self-dual after the metric changes.  Given that
\begin{eqnarray}
B_2' \wedge F_3' = f \wedge A_M \wedge (\eta + A_q) \wedge d {\cal A} \,,
\end{eqnarray}
we must have that $*_5 F_q$ becomes in the new metric,
\begin{eqnarray}
\label{HodgeStarFq}
*_5 F_q =e^{-V}  *_5' (F_q - 2 f \wedge A_M) \,,
\end{eqnarray}
and the five-forms can be written,
\begin{eqnarray}
\label{FZero}
F_5^0 &=&- 4e^{-V} {\rm vol}_{M'} + {1 \over 2} (\eta + A_q) \wedge d{\cal A} \wedge d{\cal A} \\
&&+ {1 \over 2}e^{-V} *_5' (F_q - 2 f \wedge A_M) \wedge d {\cal A}  - {1 \over 2} F_q \wedge  (\eta + A_q) \wedge d{\cal A} \,, \nonumber
\end{eqnarray}
and
\begin{eqnarray}
\label{MelvinFive}
\label{F}
F_5' &=&- 4e^{-V} {\rm vol}_{M'} + {1 \over 2} (\eta + A_q) \wedge d{\cal A} \wedge d{\cal A} \\
&&+ {1 \over 2} *_5' (F_q - 2 f \wedge A_M) \wedge d {\cal A}  - {1 \over 2} (F_q - 2 f \wedge A_M) \wedge  (\eta + A_q) \wedge d{\cal A} \,. \nonumber
\end{eqnarray}
The Bianchi identity $dF_5' = H_3' \wedge F_3'$ implies
\begin{eqnarray}
d e^{-V}  *_5' (F_q - 2 f \wedge A_M) = F_q \wedge F_q \,,
\end{eqnarray}
consistent with (\ref{FqEqn}) and (\ref{HodgeStarFq}).

\subsection{Reduction of IIB action}

The IIB action in string frame with $C_0 = 0$ is
\begin{eqnarray}
2 \kappa_{10}^2 S &=& \int \Bigg[ (* 1) e^{-2 \Phi} \Big( R + 4 (\partial \Phi)^2 \Big)  - {1 \over 2} e^{-2\Phi} H_3 \wedge * H_3 - {1 \over 2} F_3 \wedge * F_3 \\ && - {1 \over 4} F_5 \wedge * F_5 - {1 \over 2} C_4 \wedge H_3 \wedge F_3 \Bigg] \,. \nonumber
\end{eqnarray}
Here $F_5 \equiv dC_4 - C_2 \wedge H_3$ is the gauge-invariant field strength.   As usual the IIB action gives the correct equations of motion, but $F_5 = * F_5$ must be imposed only after deriving the equations.

Using the expressions (\ref{MelF}) and (\ref{MelB}) for $F_3$ and $H_3$ and dropping primes for convenience, we find
\begin{eqnarray}
* F_3 &=&e^V *_5 df \wedge (\eta^{(5)} + A_q) \wedge d {\cal A} 
+ e^{V}  *_5 df_2\wedge (\eta^{(5)} + A_q)\wedge {1 \over 8}d {\cal A}  \wedge d {\cal A} \,, \\
*H_3 &=& 
- e^{ - V} *_5 F_M \wedge {1 \over 8} d{\cal A} \wedge d{\cal A} 
 - e^V *_5 A_M \wedge (\eta^{(5)} + A_q) \wedge d {\cal A} \\ &&
- e^{ V} *_5 (A_M \wedge F_q) \wedge (\eta^{(5)} + A_q) \wedge {1 \over 8} d {\cal A} \wedge d{\cal A} \,,
\nonumber
\end{eqnarray}
The Ricci scalar is
\begin{eqnarray}
R = R^{(5)}   - 2 \partial_\mu V \partial^\mu V - 2 \nabla^2 V 
+ 24 - 4 e^{ 2V} - {1 \over 4} e^{2V} (F_q)_{\mu\nu} (F_q)^{\mu\nu} \,.
\end{eqnarray}
Using integration by parts and the fact that $d (B_2 \wedge C_2) \wedge B_2 \wedge F_3 = {1 \over 2} d(B_2 \wedge B_2 \wedge C_2 \wedge F_3)$ is a total derivative,
 the Chern-Simons term can be reexpressed as,
\begin{eqnarray}
- {1 \over 2} \int C_4 \wedge H_3 \wedge F_3 \to  {1 \over 2} \int F_5^0 \wedge B_2 \wedge F_3 \,,
\end{eqnarray}
Using the above relations,  and the ansatz (\ref{MelMet})-(\ref{MelDil}) and (\ref{FZero}) and (\ref{F}) for the RR 5-form, we perform the dimensional reduction leading to (\ref{FiveAction}). Because of the subtlety arising from the self-duality of the RR 5-form, the coefficients were checked using the mode decomposition of the $F_3$ and $H_3$ equations of motion.

\end{document}